\documentclass[aps,prl,twocolumn,letterpaper,showpacs]{revtex4-1}
\usepackage[utf8x]{inputenc}
\usepackage{graphicx,amsmath}
\usepackage{amssymb}
\usepackage{amsthm}

\begin{document}
\title{All-electric qubit control in heavy hole quantum dots via non-Abelian geometric phases}
\author{Jan C. Budich, Dietrich G. Rothe, Ewelina M. Hankiewicz, and Bj\"orn Trauzettel}
\affiliation{Institute for Theoretical Physics and Astrophysics,
University of W$\ddot{u}$rzburg, 97074 W$\ddot{u}$rzburg, Germany}
\date{\today}
\begin{abstract}
We demonstrate how non-Abelian geometric phases can be used to universally process a spin qubit in heavy hole quantum dots in the absence of magnetic fields. A time dependent electric quadrupole field is used to perform any desired single qubit operation by virtue of non-Abelian holonomy. During the proposed operations, the degeneracy of the time dependent two level system representing the qubit is not split. Since time reversal symmetry is preserved and hyperfine coupling is known to be weak in spin qubits based on heavy holes, we expect very long coherence times in the proposed setup.
\end{abstract}
\maketitle
Coherent spin control by all-electric means (without breaking time reversal symmetry (TRS)) is among the major goals of spintronics. One of the reasons why is that the presence of TRS is known to forbid several dephasing mechanisms, for example, in spin qubits \cite{LossQCQD1998}, due to the interplay of electron phonon coupling and Rashba spin orbit coupling \cite{BulaevHH2005}. In the original work by Loss and DiVincenzo \cite{LossQCQD1998}, the proposed scheme for universal quantum computing based on spin qubits in quantum dots (QDs) relied, on the one hand, on all-electric two qubit operations but, on the other hand, on single qubit operations based on magnetic fields or ferromagnetic auxiliary devices that both break TRS. A few years later, electric-dipole-induced spin resonance (EDSR) has been proposed \cite{GolovachEDSR} and experimentally realized \cite{NowackEDSR} as a way to process spins electrically in presence of a static magnetic field which is still breaking TRS. Rather recently, it has been theoretically shown that in spin qubits based on carbon nanotube QDs it is indeed possible to accomplish all-electric single qubit operations using EDSR \cite{Bulaev2008,Klinovaja2011}. This is true because the specific spin orbit interaction in carbon nanotubes provides a way to split spin up and spin down states in the absence of magnetic fields. However, spin qubits based on carbon nanotubes face other problems and it is fair to say that all host materials for spin qubits have advantages and disadvantages.

\begin{figure}
\centering
\includegraphics[width=0.85\columnwidth]{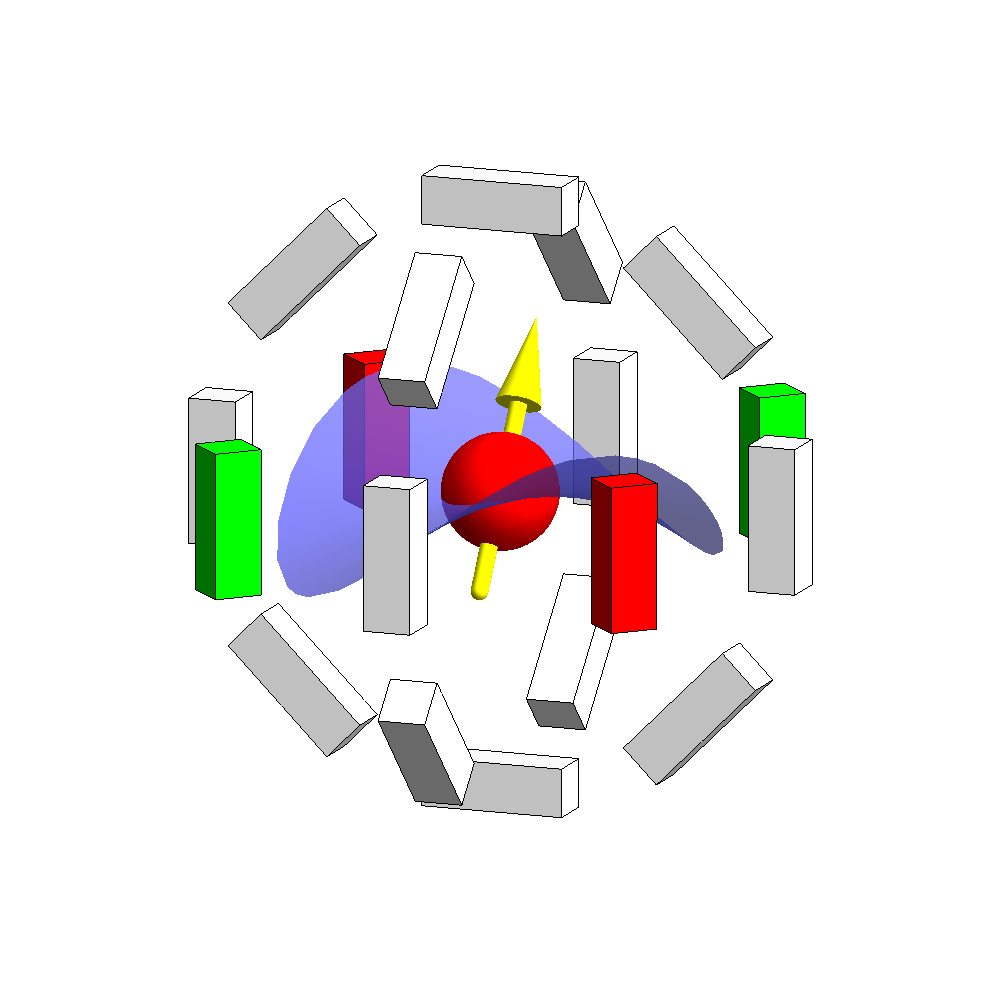}
\caption{(Color online) Schematic of single particle (red ball) with a HH (pseudo)spin (yellow arrow) in a $J= \frac{3}{2}$~valence band QD. The three-dimensional QD is surrounded by 18 gates that allow to generate an electrostatic potential with quadrupole symmetry in any direction in space. The red and green colors on the gates should visualize applied electrostatic potentials with opposite sign that give rise to the quadrupole field shown in light blue.}
\label{fig:setup}
\end{figure}

In this work, we are interested in spin qubits based on heavy hole (HH) QDs. We show how universal single qubit operations can be performed by all-electric means in the framework of holonomic quantum computing \cite{Zanardi1999} in these systems. The adiabatic evolution in presence of a time dependent electric quadrupole field is employed to control the HH qubit (see Fig. \ref{fig:setup} for a schematic). For our purposes, HH spin qubits (composed of $J=\frac{3}{2}$ states) are the simplest two level system that can be manipulated in the desired way. However, HH spin qubits are, of course, a very active research area by itself beyond holonomic quantum computing. Two reasons why HH QDs are promising and interesting candidates for spin qubits are, for instance, the advanced level of optical control \cite{Gerardot2008,Eble2009,Brunner2009,YamamotoHH2011} and the predicted long coherence times \cite{FischerHH2010}.

The topological properties of TRS preserving half integer spin systems have been analyzed in a series of seminal papers by Avron and coworkers \cite{Avron1988,AvronQuadrupole1989}. The case $J=\frac{3}{2}$~is of particular interest both from a theoretical and from a more applied point of view. From the theoretical side, all TRS preserving gapped Hamiltonians are unitarily related due to an $\rm{SO}(5)$~symmetry \cite{Avron1988,AvronQuadrupole1989} giving rise to an $\rm{SO}(5)$~Clifford algebra \cite{AvronQuadrupole1989,Demler1999} which allows for a simple analytical calculation of the adiabatic time evolution and with that the geometric phase. From the experimental side, the $J=\frac{3}{2}$~system is naturally realized in the $p$-like valence band of many semiconductors where spin orbit coupling isolates the $J=\frac{3}{2}$~states from the so called split-off band. Interestingly, the fingerprints of $SU(2)$ non-Abelian geometric phases \cite{Kato1950,WilczekZee1984} could be also identified in several transport properties of this class of semiconducting materials \cite{Arovas1998,Murakami2004}.

The pioneering idea of using non-Abelian holonomy to perform quantum computing tasks in the $J=\frac{3}{2}$~system is due to Bernevig and Zhang \cite{BernevigHolQC2005} who proposed the electric Stark effect to process valence band impurities in III-V semiconductors. The idea works for the light hole (LH) subspace of the $J=\frac{3}{2}$~Hilbert space. However, the resulting holonomy is Abelian on the HH subspace \cite{ZeeNuclearBerry1988} so that the electric Stark-effect cannot be used to process HH qubits. In contrast, the electric quadrupole fields employed in our proposal allow for full adiabatic control over the entire $J=\frac{3}{2}$~Hilbert space. This is a consequence of the topologically nontrivial structure of the accessible parameter space which becomes manifest in the nontrivial second Chern number of the associated $SU(2)$ gauge theory over the space of quadrupole tensors \cite{Avron1988}.

Recently, holonomic quantum computing due to tunable spin orbit coupling with electron spins in spatially transported quantum dots has been suggested \cite{ ZarandPRB2007, GolovachLoss2010} but this is very demanding from an experimental point of view. Our idea is conceptually much simpler. We derive below the time dependent electric quadrupole field that realizes any desired single qubit operation
\begin{align}
\mathcal U(\hat n, \varphi)=\exp\left({i\varphi \frac{\hat n \vec \sigma}{2}}\right)
\label{eqn:su2gen}
\end{align}
on the HH spin qubit. Here, $\hat n$~is a unit vector representing the rotation axis, $\varphi$~is the angle of the rotation, and $\vec \sigma$~denotes the vector of Pauli matrices acting on the qubit space. Furthermore, we give an estimate of the adiabatic time scale which determines the maximum operating frequency of single qubit gates showing that the physics we describe is experimentally feasible. Finally, we discuss the influence of several imperfections, which might be present in an experimental setup, on the working precision of our proposal.

The non-Abelian geometric phase, occurring in a degenerate subspace after a cyclic evolution, is readily expressed once the time-dependent projection $P(t)$~onto this degenerate subspace is known. The generator of the adiabatic evolution then reads \cite{Kato1950}
\begin{align}
\mathcal A\left(\frac{d}{dt}\right) = -\left[\frac{d P(t)}{dt},P(t)\right] .
\end{align}
On the basis of this generator, the non-Abelian geometric phase \cite{WilczekZee1984} associated with a loop $\gamma$~in parameter space is given by the holonomy
\begin{align}
U_\gamma = \mathcal T \text{e}^{\int_\gamma \mathcal A},
\label{eqn:hol}
\end{align}
where $\mathcal T$~denotes time-ordering. For the Hilbert space of a $J=\frac{3}{2}$~particle in presence of TRS this holonomy is readily calculated analytically as we explicitly demonstrate below.

The Hamiltonian of a spin $\frac{3}{2}$~particle coupled to an electric quadrupole field can be written as \cite{Avron1988}
\begin{align}
H(\mathcal Q)=  J_i\mathcal Q^{ij}J_j,
\label{eqn:quadham}
\end{align}
where $J$~is the angular momentum operator and $\mathcal Q$~is the quadrupole tensor of the applied field. We put $\hbar = 1$ in the following. $\mathcal Q$~is a real, symmetric, traceless matrix. The space of such matrices is five dimensional. An orthonormal basis of this space is given by the matrices $\left\{Q_\mu\right\}_\mu,~\mu=0,\ldots,4$ (see Appendix for more details), which satisfy $\frac{3}{2}\text{Tr}\left\{Q_\mu Q_\nu\right\}=\delta_{\mu\nu}$. A general quadrupole field is then of the form $x^\mu Q_\mu$~and the associated Hamiltonian reads
\begin{align}
H(\mathcal Q)=H(x^\mu Q_\mu)=x^\mu J_i Q_\mu^{ij}J_j\equiv  x^\mu \Gamma_\mu,
\end{align}
where the basis Hamiltonians $\Gamma_\mu \equiv J_i Q_\mu^{ij}J_j$ obey the $\rm{SO}(5)$~Clifford algebra \cite{AvronQuadrupole1989,Demler1999}
\begin{align}
\left\{\Gamma_\mu,\Gamma_\nu\right\}=2\delta_{\mu\nu}.
\end{align}
As far as the geometric phase associated with a cycle in this parameter space is concerned, we can confine our interest to quadrupole fields of constant strength, say $\lvert\mathbf{ x}\rvert = 1$. (This is justified because the quadrupole energy is the only energy scale of the problem.) Note that the experimentally relevant scale of $ \lvert\mathbf{ x}\rvert$~defines the splitting between the two Kramers pairs and therefore the adiabatic operating frequencies of the proposed setup. Due to the mentioned $\rm{SO}(5)$~symmetry in the system \cite{Avron1988}, all possible quadrupole Hamiltonians $H(\mathcal Q)$~are unitarily related by a $\rm{Spin}(5)$~representation of this $\rm{SO}(5)$~symmetry. The ten generators of this symmetry group of our family of Hamiltonians are given by \cite{AvronQuadrupole1989}
\begin{align}
\left\{V_i\right\}_i=\left\{\frac{1}{2}\left[\Gamma_\alpha,\Gamma_\beta\right]=\Gamma_\alpha \Gamma_\beta\right\}_{\alpha <\beta},~i\in{0,\ldots,9} ,
\end{align}
where $\alpha,\beta \in 0, \dots, 4$. A cyclic time evolution $t\mapsto H(t)$~starting from $H(t=0)= \Gamma_0$~is then given by a $2\pi$~$\rm{SO}(5)$~rotation on the space of quadrupole fields which is uniquely associated with a $2\pi$~$\rm{Spin}(5)$~rotation
\begin{align}
t\mapsto H(t)=\text{e}^{t\frac{\hat a \vec V}{2}} \Gamma_0\text{e}^{-t\frac{\hat a \vec V}{2}},~t\in\left[0,2\pi\right],
\end{align}
in Hilbert space, where $\hat a$~is a ten-component unit vector specifying the direction of the rotation in the Lie algebra of $\rm{SO}(5)$. We call
\begin{align}
P^\pm_0=\frac{1}{2}(1\pm \Gamma_0)
\end{align}
the projector on the Kramers pair with eigenvalue $\pm \lvert \mathbf{x}\rvert$. In fact, due to our choice of the initial Hamiltonian $P^\pm_0$~concurs with the projection on the HH/LH subspaces. Starting with a HH state $\lvert \psi(0)\rangle$~satisfying $P^+_0\lvert \psi(0)\rangle=\lvert \psi(0)\rangle$~the adiabatic time evolution $U(t)$ can be conveniently expressed as \cite{Simon1983,WilczekZee1984,AvronQuadrupole1989}
\begin{align}
&U(t)=\lim_{n\rightarrow\infty} U_n(t)~\text{with}\nonumber\\
&U_n(t)=P^+(t)P^+\left(\frac{(n-1)t}{n}\right)\cdots P^+\left(\frac{2t}{n}\right)P^+\left(\frac{t}{n}\right)P^+_0,
\label{eqn:adiabatic}
\end{align}
where the time dependent projector on the Kramers pair with positive eigenvalue is given by $P^+(t)=\text{e}^{t\frac{\hat a \vec V}{2}} P^+_0\text{e}^{-t\frac{\hat a \vec V}{2}}$. Along any such loop $\gamma$~in parameter space the adiabatic evolution is readily computed analytically to yield \cite{AvronQuadrupole1989}
\begin{align}
U(t)=\text{e}^{t\frac{\hat a \vec V}{2}}\text{e}^{-tP_0^+\frac{\hat a \vec V}{2}P_0^+}.
\label{eqn:adevolution}
\end{align}
The first factor gives $\text{e}^{2\pi\frac{\hat a \vec V}{2}}=-1$~once the loop is completed. The second factor at $t=2\pi$~ defines an $\rm{SU}(2)$~transformation on the HH subspace which is the desired holonomy $U_\gamma$ (see Eq. (\ref{eqn:hol})) up to a sign. Note that the holonomy associated with a loop $\gamma$~is a purely geometrical object. It does not depend on parameterization, i.e. on the time-dependent velocity with which the electric field is ramped, as long as the adiabatic approximation is justified.\\
We now explicitly construct the direction $\hat a$~needed to obtain any holonomy as parameterized in Eq. (\ref{eqn:su2gen}). The angle and axis of the rotation can be tuned using the relations
\begin{align}
P^+_0\Gamma_0\Gamma_\mu P^+_0=0,~\mu\ne 0,
\label{eqn:orthpart}
\end{align}
as well as
\begin{align}
&P^+_0\Gamma_4\Gamma_1 P^+_0=i\sigma_x,~P^+_0\Gamma_1\Gamma_3 P^+_0=i\sigma_y,\nonumber\\
& P^+_0\Gamma_1\Gamma_2 P^+_0=i\sigma_z,
\label{eqn:parpart}
\end{align}
where $\sigma_i$~ are the Pauli matrices on the HH subspace with eigenvalue $+\lvert\mathbf{x}\rvert$. To see this, let us restrict ourselves to the four generators $\Gamma_1\Gamma_\mu,~\mu\ne 1$~and label them $V_0 = \Gamma_0\Gamma_1$, $V_1 = \Gamma_4\Gamma_1$, $V_2 = \Gamma_1\Gamma_3$, $V_3 = \Gamma_1\Gamma_2$. With this restriction $\hat a$~only has the nonvanishing components $a_0,a_1,a_2,a_3$~satisfying $\sum_{i=0}^3a_i^2=1$. Using Eqs. (\ref{eqn:adevolution}-\ref{eqn:parpart}) we get by comparison to Eq. (\ref{eqn:su2gen})
\begin{align}
&\varphi = 2\pi\left(1- \sqrt{\sum_{i\ne 0} a_i^2}\right) =2\pi\left(1-\sqrt{1-a_0^2}\right)\in \left[0,2\pi\right],\nonumber\\
&\hat n  = \frac{(a_1,a_2,a_3)}{\lvert (a_1,a_2,a_3)\rvert}.
\label{eqn:antrans}
\end{align}
Next, we translate the loop associated with the direction $\hat a$~into a time dependent quadrupole field. To do so, we write the time dependent Hamiltonian $H(t)=x^\mu(t)\Gamma_\mu$~in two different ways
\begin{align}
H(t)= \text{e}^{t \frac{\hat a\vec V}{2}}x^\mu(0) \Gamma_\mu\text{e}^{-t \frac{\hat a\vec V}{2}} =\left(\text{e}^{t\hat a\vec W}\mathbf{x}(0)\right)^\mu \Gamma_\mu,
\end{align}
where $\vec W$~represents the $\rm{SO}(5)$~generators in the defining representation acting on the $\mathbb R^5$~vector $\mathbf{x}$. Spelling the latter equation out for infinitesimal transformations and using the independence of the different $\Gamma_\mu$, i.e. $\frac{1}{4}\text{Tr}\left\{\Gamma_\mu \Gamma_\nu\right\}=\delta_{\mu\nu}$, we obtain the desired $\rm{SO}(5)$ generators $W_0,\ldots,W_3$ (see Appendix) associated with the $\rm{Spin}(5)$~generators $V_0,\ldots,V_3$. Now, we can define the time dependent quadrupole field associated with the loop in direction $\hat a$:
\begin{align}
\mathcal Q(t)= x^\mu(t)Q_\mu = \left(\text{e}^{t\hat a\vec W}\mathbf{x}(0)\right)^\mu Q_\mu,\quad t\in \left[0,2\pi\right]
\end{align}
which needs to be experimentally applied to perform the desired single qubit operation.

\begin{figure}
 \includegraphics[width=\columnwidth]{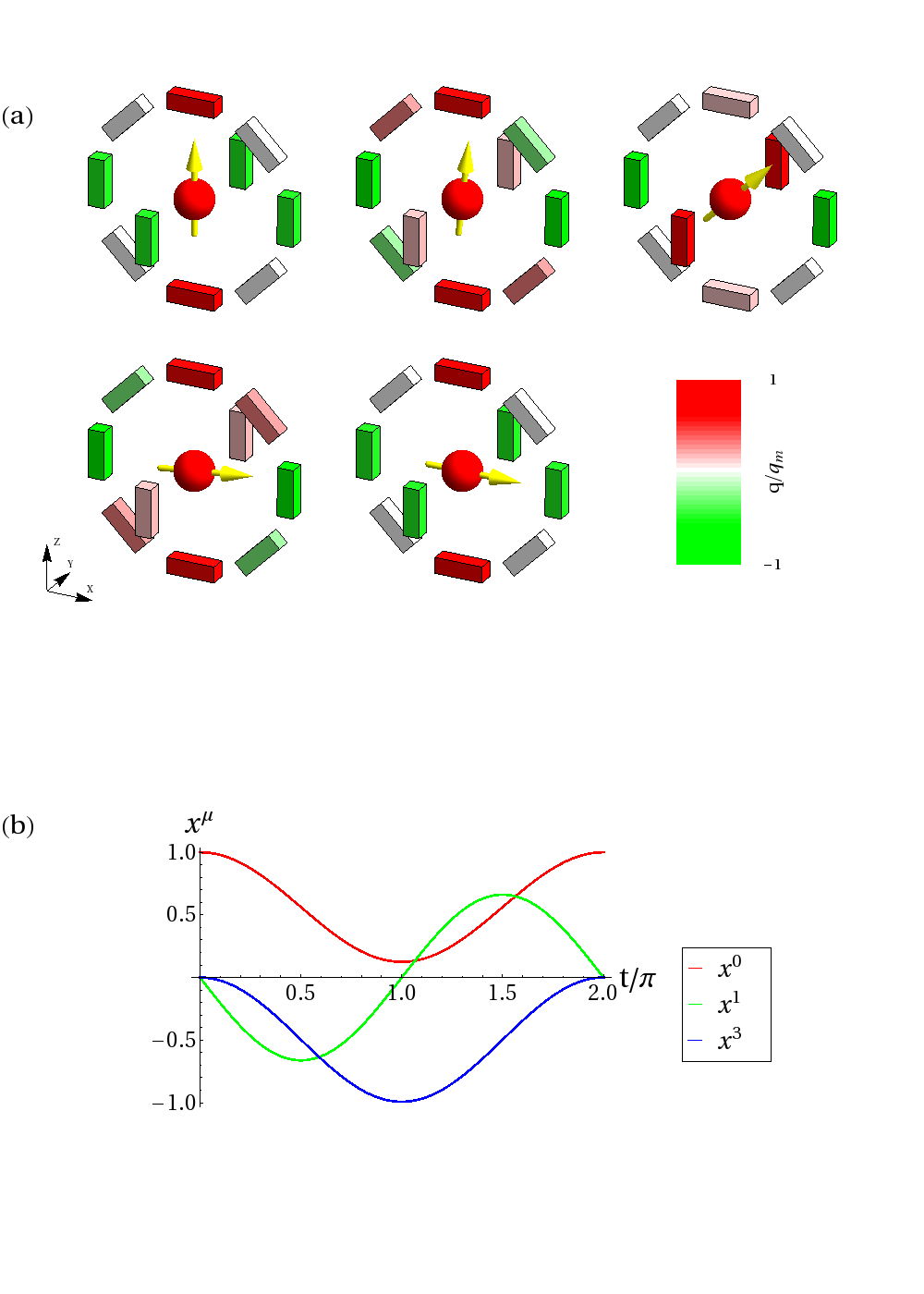}

 \caption{(Color online)
 (a) A 10-gate setup realizing the operation $\mathcal U(-\hat e_y,\frac{\pi}{2})$~ on the HH spin (yellow arrow). The colors of the schematic gates visualize their time-dependent charge during the loop operation, at times from left to right and top to bottom, $t=0,\frac{\pi}{2},\pi,\frac{3\pi}{2}$, and $2\pi$. All charges are normalized to the charge $q_{m}$ of the topmost gate at $t=0$.
 (b) Time-dependence of non-zero components of $\mathbf{x}$~during the operation $\mathcal U(-\hat e_y,\frac{\pi}{2})$.
 }
\label{fig:example}
\end{figure}

Let us give a concrete example. If we were to rotate the HH spin from pointing in $z$-direction to the $x$-direction this would correspond to the operation $\mathcal U(-\hat e_y,\frac{\pi}{2})=\frac{1}{\sqrt{2}}\begin{pmatrix} 1&{-1}\\{1}&1\end{pmatrix}$~which is associated with the quadrupole field
\begin{align}
Q(t)= \left(\text{e}^{t\left(\frac{\sqrt{7}}{4}W_0-\frac{3}{4}W_2\right)}\mathbf{e_0}\right)^\mu Q_\mu,\quad t\in \left[0,2\pi\right],
\end{align}
i.e. $\hat a = (a_0,a_1,a_2,a_3)=(\frac{\sqrt{7}}{4},0,-\frac{3}{4},0)$~and $\mathbf{x}(t=0)=\mathbf{e_0}=(1,0,0,0,0)$~in the language of our general analysis. Indeed plugging this choice of $\hat a$ into Eq. (\ref{eqn:antrans}) yields $\hat n= -\hat e_y,~\varphi =\frac{\pi}{2}$. A stroboscopic illustration of a possible electrostatic gating scheme realizing this time-dependent quadrupole field is shown in Fig.~\ref{fig:example}. For this particular example, we only need 10 of the 18 gates illustrated in Fig. \ref{fig:setup}. To perform an arbitrary $\rm{SU}(2)$ transformation 14 of these 18 gates are needed. We could drop, for instance, the four gates that are colored in red and green in Fig. \ref{fig:setup} and still be able to perform any desired single qubit rotation on the HH subspace.

Up to now, the energy scale $\Delta E = \lvert \mathcal Q\rvert=\lvert \mathbf{x}\rvert$ (see Eq.~(\ref{eqn:quadham})) has been treated as a free parameter. To show that this scale is amenable to state of the art experiments on GaAs quantum dots, we give a numerical estimate for $\Delta E$. To do so, we calculate the HH-LH splitting $\Delta E$~associated with an electrostatic potential $e\Phi_4(\vec r) = \lambda \vec r^T\mathcal Q \vec r$~with quadrupole symmetry on the basis of a Luttinger four-band model for the valence bands of a GaAs/AlGaAs quantum well \cite{Adreani1987,ChuangPRB}. Here, $\vec r$~denotes the real space position vector and the QDs are modelled by a parabolic lateral confinement potential defining the dots on a typical length scale of $50 {\rm nm}$. The strength of the potential is determined by the constant $\lambda$. For a realistic quadrupole potential $e\Phi_4 \sim 50{\rm meV}$ at a distance $r\sim50{\rm nm}$ away from the center of the dot, we obtain a splitting of $\Delta E = 0.57 {\rm meV}$, which corresponds to a temperature of $6.6$~K and an adiabatic frequency of $\omega = 0.87{\rm THz}$~respectively (see Appendix). Therefore, it is easily possible to stay below this frequency such that the adiabatic evolution is justified and at the same time complete the loop much faster than typical dephasing times in HH spin qubits. ($T_2$ of the order of $\mu$s has been measured in Ref.~\cite{YamamotoHH2011}.)

In real experiments, there will not only be the (wanted) HH-LH splitting $\Delta E$ induced by the quadrupole field but also an (unwanted) HH-LH splitting $\Delta E_0$ induced by confinement. For our purposes, the former should be much larger than the latter. We estimate in the Appendix that often times it is the other way round, i.e.~$\Delta E_0$ is much larger than $\Delta E$ which is a true problem for our proposal. However, by applying mechanical strain, the splitting of the individual quadruplet subbands on the quantum dot can be engineered significantly \cite{Adreani1987, ChuangBook} (see also Appendix). For the parameters used in our model, the confinement induced splitting $\Delta E_0$ can then be realistically tuned below our estimated value of $\Delta E = 0.57 {\rm meV}$ (see Appendix). Hence, strain engineering of the QD is needed to guarantee a reliable performance of our setup. Additionally, we note that our proposal is robust against unwanted residual dipole fields, deviations from a quadrupole potential with only $l=2$ contributions, and deviations from a quadratic confinement potential. The influence of these perturbations on $\Delta E$ are carefully analyzed in the Appendix and shown to be harmless.

In summary, we have demonstrated that an electric quadrupole field can be used to fully control a HH qubit without breaking TRS. The adiabatic time scale of our proposal is determined by the field induced splitting $\Delta E$ between the two Kramers pairs, which we have estimated for GaAs QDs to be on the order of $0.57 {\rm meV}$. The maximum operating frequency of the device should be significantly below this energy scale to justify the adiabatic approximation which is understood throughout our analysis. Confinement induced splitting between the two Kramers pairs in the $J=\frac{3}{2}$~quadruplet of levels at the relevant energy in the HH QD impinges on the efficiency of the geometric control over the qubit. The scale of this splitting for a given quadruplet can be tuned/reduced by applying strain. We note that exact control over the qubit is still possible as long as the quadrupole energy gap is larger than the confinement induced splitting. Our proposal is not limited to HH quantum dots in GaAs quantum wells, but can in principle also be employed to process trapped spin $\frac{3}{2}$~ions or HH-like valence band impurities by means of a quadrupole field. The presence of TRS in combination with suppressed hyperfine coupling in HH systems renders our proposal less prone to decoherence than nonadiabatic processing schemes relying on the presence of a Zeeman splitting due to an external magnetic field. Two-qubit gates can be performed by virtue of electrostatic gates as proposed in Ref.\cite{LossQCQD1998}. All-electric spin pumping and spin filtering techniques respectively \cite{ZarandSpinPumping} can be used to perform initialization and readout tasks on the quantum dots. Hence, our proposal in principle allows for TRS preserving universal quantum computing.

We acknowledge helpful discussions with Jan Jacob, Patrik Recher, and Ronny Thomale as well as financial support by the DFG (grant HA5893/1-2 and the JST-DFG research unit ``Topotronics'') and the ESF (RNP QSpiCE).

\section{Appendix}
\label{sec:som}
\subsection*{Quadrupole Hamiltonians and SO$(5)$~symmetry}
In this section we review some properties of the space of time reversal invariant traceless spin $\frac{3}{2}$~Hamiltonians referred to in the main text.
The familiy of quadrupole Hamiltonians
\begin{align*}
H(\mathcal Q)= H(x^\mu Q_\mu)=x^\mu J_i Q_\mu^{ij}J_j\equiv  x^\mu \Gamma_\mu
\end{align*}
is parameterized by the space of quadrupole tensors, i.e. real, symmetric, and traceless $3\times 3$~matrices.
An orthogonal basis of this five dimensional space is given by $\left\{Q_\mu\right\}_\mu,~\mu=0,\ldots,4$, with
\begin{align*}
 &Q_0 = \frac{1}{3}\begin{pmatrix} {-1} & 0&0 \\ 0& {-1}&0\\0&0&2 \end{pmatrix},~Q_1 = \frac{1}{\sqrt{3}}\begin{pmatrix} 0 & 0&1 \\ 0& 0&0\\1&0&0 \end{pmatrix},\nonumber\\
&Q_2 = \frac{1}{\sqrt{3}}\begin{pmatrix} 0 & 0&0 \\ 0& 0&1\\0&1&0 \end{pmatrix},~
Q_3 = \frac{1}{\sqrt{3}}\begin{pmatrix} 1 & 0&0 \\ 0& {-1}&0\\0&0&0 \end{pmatrix},\nonumber\\
&Q_4 = \frac{1}{\sqrt{3}}\begin{pmatrix} 0 & 1&0 \\ 1& 0&0\\0&0&0 \end{pmatrix}
\end{align*}
which satisfy the normalization condition
\begin{align*}
\frac{3}{2}\rm{Tr}\left\{Q_\mu Q_\nu\right\}=\delta_{\mu\nu}.
\end{align*}
The basis Hamiltonians $\Gamma_\mu = J_i Q_\mu^{ij}J_j$ obey the $\rm{SO}(5)$~Clifford algebra
\begin{align*}
\left\{\Gamma_\mu,\Gamma_\nu\right\}=2\delta_{\mu\nu}.
\end{align*}
The generators of the $\rm{Spin}(5)$~group action on the space of quadrupole Hamiltonians are readily expressed in terms of the basis Hamiltonians as
\begin{align*}
&\left\{V_i\right\}_i=\left\{\frac{1}{2}\left[\Gamma_\alpha,\Gamma_\beta\right]=\Gamma_\alpha \Gamma_\beta\right\}_{\alpha <\beta};\nonumber\\
&i\in{0,\ldots,9};~\alpha,\beta\in{0,\ldots,4}.
\end{align*}
Spelling out the orbit $t\mapsto H(t)$~ of this group action in two different ways,
\begin{align*}
H(t)= \text{e}^{t \frac{\hat a\vec V}{2}}x^\mu(0) \Gamma_\mu\text{e}^{-t \frac{\hat a\vec V}{2}} =\left(\text{e}^{t\hat a\vec W}\mathbf{x}(0)\right)^\mu \Gamma_\mu,
\end{align*}
the correspondence between the $\rm{Spin}(5)$~action on the space of Hamiltonians and the $\rm{SO}(5)$~action on the space $\mathbb R^5 \setminus\left\{0\right\}$~of parameter vectors $\mathbf{x}$~becomes manifest.
The $SO(5)$~defining representation $W_0\ldots W_3$ of the $\rm{Spin}(5)$~generators $V_0\ldots V_3$~defined in the main text reads
\begin{align*}
W_0 = \begin{pmatrix} 0 & 1&0&0&0 \\ {-1}&0&0&0&0\\0&0&0&0&0\\0&0&0&0&0\\0&0&0&0&0 \end{pmatrix},~
W_1 = \begin{pmatrix} 0 &0&0&0&0 \\ 0&0&0&0&{-1}\\0&0&0&0&0\\0&0&0&0&0\\0&1&0&0&0 \end{pmatrix},\\
W_2 = \begin{pmatrix} 0 & 0&0&0&0 \\ 0&0&0&1&0\\0&0&0&0&0\\0&{-1}&0&0&0\\0&0&0&0&0 \end{pmatrix},~
W_3 = \begin{pmatrix} 0 & 0&0&0&0 \\ 0&0&1&0&0\\0&{-1}&0&0&0\\0&0&0&0&0\\0&0&0&0&0 \end{pmatrix}.
\end{align*}

\subsection*{Quadrupole induced HH/LH splitting in strained GaAs quantum dots}
In this section, we give a quantitative estimate of the heavy hole (HH)/light hole (LH) splitting induced by an electric quadrupole field on strained GaAs quantum dots. We model a quantum dot using the effective 2D Hamiltonian of a [001] quantum well \cite{Adreani1987} (QW) and add some parabolic confinement $\Phi_1 (x,y)$.
This reduces the symmetry to $D_{2d}$  
and therefore, even without a quadrupole potential, we expect a HH/LH splitting $\Delta E_0$.
We account for this by extending the Hamiltonian $H(\mathcal Q)=J^T\mathcal Q J$~to
\begin{equation*}
 H' =H(\mathcal Q) + \frac{\Delta E_0}{2} \tau_z
\end{equation*}
 with $\tau_z = \mathrm{diag}(1,-1,-1,1)$ and the Hamiltonian is written in the basis $\left\{\left|\frac{3}{2},\frac{3}{2}\right\rangle,\left|\frac{3}{2},\frac{1}{2}\right\rangle,\left|\frac{3}{2},-\frac{1}{2}\right\rangle,\left|\frac{3}{2},-\frac{3}{2}\right\rangle \right\}$. Without loss of generality,
  we use a quadrupole potential $e \Phi_4 = \frac{1}{R^2} \vec r^T \mathcal Q\vec r$ associated with the quadrupole tensor of four
Coulomb charges $\pm q$ at equal radius $R$ in the $(x,y)$ plane (corresponding to $\lambda = \frac{1}{R^2}$ in the main text),
\begin{equation*}
\mathcal{Q} =  \frac{1}{4 \pi \epsilon}  \frac{6 e q}{R} \left( \begin{array}{ccc} 1 & 0 & 0 \\ 0 & -1 & 0 \\ 0 & 0 & 0 \end{array}\right).
\end{equation*}

Whereas the spectrum of $H$ is $E = \pm |\mathbf{x}|$, where $\mathbf{x}$ is a 5-component vector
defined by the expansion $\mathcal{Q} = x^\mu Q_\mu$,
the spectrum of $H'$~simplifies for our choice of the quadrupole potential to
\begin{equation}
\label{dispeff}
 E = \pm \frac{1}{2} \sqrt{\Delta E_0^2 + 4 \lvert \mathbf{x}\rvert^2}.
\end{equation}
We will use this relation to fit $\lvert \mathbf{x}\rvert$~as a function of the strength of the quadrupole potential.
To obtain an effective Hamiltonian for the QW, we first solve the
envelope function $\vec{f}(z)$~where $z$~is the direction of growth.
In general, the 4-component envelope function $\vec{f}(z)$~depends on $k_\|=(k_x,k_y)$.
We simplify the problem by  performing a  k $\cdot$ p calculation with expansion of $k_\|$ around the $\Gamma$ point.
The Luttinger Hamiltonian $H_L(k_\|=0)$~is diagonal and for the $i$th component $f_i$ of $\vec{f}$ we find
\begin{equation*}
\left(k_z \frac{1}{2m_i(z)} k_z + V(z)  \right) f_i(z) = E_i f_i(z).
\end{equation*}
Here, $m_i(z)$ is the material dependent bulk effective mass, which is $m_{B,i}$ for the barrier and
$m_{W,i}$ for the material of the well and band dependent (index $i$). Furthermore, $V(z)=V_B$ in the barrier and zero otherwise. We use the symmetric ansatz
\begin{equation*}
 f_i(z) = \left\{ \begin{array}{cc} A_i e^{\xi_i (z+W/2)} & z < -W/2, \\
                 B_i \cos(k_i z)                            & -W/2 \le z \le W/2, \\
                 A_i e^{-\xi_i (z - W/2)}               & z > W/2,
                \end{array} \right.
\end{equation*}
where $W=60$nm is the QW width,
$k_i = \sqrt{2 m_{W,i} E_i}$ and $\xi_i = \sqrt{2 m_{B,i} (V_B -E_i)}$. Continuity of $f_i(z)$ and $m_i(z) f_i'(z)$
give the secular equation
\begin{equation*}
\sqrt{1 - \frac{1}{\tilde{k_i}^2}} = \left(\frac{m_{B,i}}{m_{W,i}}\right)^{3/2} \tan \left( \tilde{k_i} W \sqrt{\frac{m_{W,i} V_B}{2}} \right)
\end{equation*}
with $\tilde{k_i} \sqrt{2 m_{W,i} V_B} = k_i$.

The Luttinger Hamiltonian for $\Gamma_8$ bands including corrections due to strain reads
\begin{equation*}
 H_L  =  - \left( \begin{array}{cccc} P + Q & -S & R & 0 \\
                     -S^\dagger & P - Q & 0 & R      \\
                     R^\dagger &  0 & P-Q & S   \\
                     0  & R^\dagger & S^\dagger & P + Q  \end{array} \right)
\end{equation*}
written in the basis $\left\{\left|\frac{3}{2},\frac{3}{2}\right\rangle,\left|\frac{3}{2},\frac{1}{2}\right\rangle,\left|\frac{3}{2},-\frac{1}{2}\right\rangle,\left|\frac{3}{2},-\frac{3}{2}\right\rangle \right\}$.
The strain tensor $\epsilon_{ij}$ gives the displacement of an atom at unit vector $\hat{i}$ along unit vector $\hat{j}$.
We consider only uniaxial strain with $\epsilon_{xx} = \epsilon_{yy} \ne \epsilon_{zz}$ and $\epsilon_{xy}=\epsilon_{xz}=\epsilon_{yz}=0$.
Then, only $P$ and $Q$ include corrections due to strain:
\begin{eqnarray*}
 && P = t_0 \gamma_1 (k_x^2 + k_y^2)  + t_0 k_z \gamma_1 k_z + P_{\epsilon},
 \\&& Q = t_0 \gamma_2 (k_x^2 + k_y^2)  - 2 t_0 k_z \gamma_2 k_z + Q_{\epsilon},
 \\&&  R = t_0 \sqrt{3} (-\gamma_2 (k_x^2 - k_y^2) + 2 i \gamma_3 k_x k_y ),
 \\&&  S = t_0 \sqrt{3} (k_x - i k_y) \{\gamma_3, k_z\}
\end{eqnarray*}
with $t_0 = \frac{1}{2 m_0}$ and
\begin{eqnarray*}
 && P_{\epsilon} = - a_v (\epsilon_{xx} + \epsilon_{yy} + \epsilon_{zz} ),
 \\ && Q_{\epsilon} = -\frac{b}{2} (\epsilon_{xx} + \epsilon_{yy} - 2 \epsilon_{zz} ).
\end{eqnarray*}
The GaAs/AlAs lattice constants are almost the same ($5.65 {\text \AA}$ vs. $5.66 {\text \AA}$). This is desirable because
one needs rather wide quantum wells and intends to avoid uncontrolled relaxation of strain. Here, we assume additional strain due to external pressure $\tau_{zz}$ which can be expressed in terms of the stiffness tensor $C$~relating strain and stress. The condition of no transversal stress
$\tau_{xx} = \tau_{yy} = 0$ gives
\begin{eqnarray*}
&& \epsilon_{xx} = \epsilon_{yy} = \frac{-C_{12}}{C_{11}^2 + C_{11} C_{12} - 2 C_{12}^2} \tau_{zz}
\\&& \epsilon_{zz} = \frac{C_{11} + C_{12}}{C_{11}^2 + C_{11} C_{12} - 2 C_{12}^2} \tau_{zz} ,
\end{eqnarray*}
where $C_{11} = 11.88~10^5{\rm bar},~C_{12} = 5.38 ~10^5{\rm bar}$. \cite{ChuangBook} We take the same values for barrier and QW for the
deformation potentials, $a_v= 1.16 \text{eV} $ and $b=-1.7 \text{eV}$. \cite{ChuangBook}
 The parameter $\zeta := Q_\epsilon$ will be used as strain control. A pressure of $1 {\rm kbar}$~gives $\zeta = 2.61 \text{meV}$.
$P_\epsilon$ is an unimportant overall energy shift.\\

The effective QW Hamiltonian is obtained by integration over envelope functions $f_\alpha(z)$ of the lowest LH and HH QW subbands,
\begin{equation*}
 H^{QW}_{\alpha \beta} = \int \mathrm{d} z \, f_\alpha^\dagger(z) H_L f_\beta(z)
\end{equation*}
Contributions of higher subbands give quantitative, but not qualitative changes of our estimates, since strain gives a diagonal correction to $H^{QW}$ and
can be used to tune $\Delta E_0$. Together with the in-plane potentials $\Phi_i$, $H^{QW}$ gives our QD model which is numerically diagonalized.

For a quantitative estimate of $|\mathbf{x}|$, we use the same parameters as Ref. \onlinecite{Adreani1987}:
For GaAs, $\gamma_1 = 6.85$, $\gamma_2 = 2.1$, $\gamma_3 = 2.9$.
For AlAs, $\gamma_1 = 3.45$, $\gamma_2 = 0.68$, $\gamma_3 = 1.29$.
The barrier material is $\text{Al}_{1-\nu}\text{Ga}_{\nu}\text{As}$ with $\nu=0.21$ and the Luttinger parameters are obtained by linear interpolation.
The bulk gap difference is $\Delta E_{g} = (1.04 \nu + 0.47 \nu^2) \text{eV} = 0.239 \text{eV}$.
We assume that the valence band shift from well to barrier is $-0.4 \Delta E_{g}$.

For the in-plane confinement, we use $e\Phi_1 = -0.15 \text{eV} (r/R_{max})^2$ where $r^2 = x^2 + y^2$. $e\Phi_1(R_{max})$ should not exceed $\Delta E_g$.
$\Phi_1$ is discretized on a 
lattice corresponding to $L$=100 nm side length, so $R_{max} = 50nm$.
By choosing $R_{max}$ and $W$ comparable, we intend to have
about the same level spacing due to in-plane and QW confinement. Then, the confinement comes closer to the ideal, fully rotationally
symmetric confinement.

With this geometry, a value of $e\Phi_{i}(R_{max}) = -1~\text{eV}$ gives a field strength of 40 meV/nm at $R_{max}$.
\begin{figure}
 \includegraphics[width=\columnwidth]{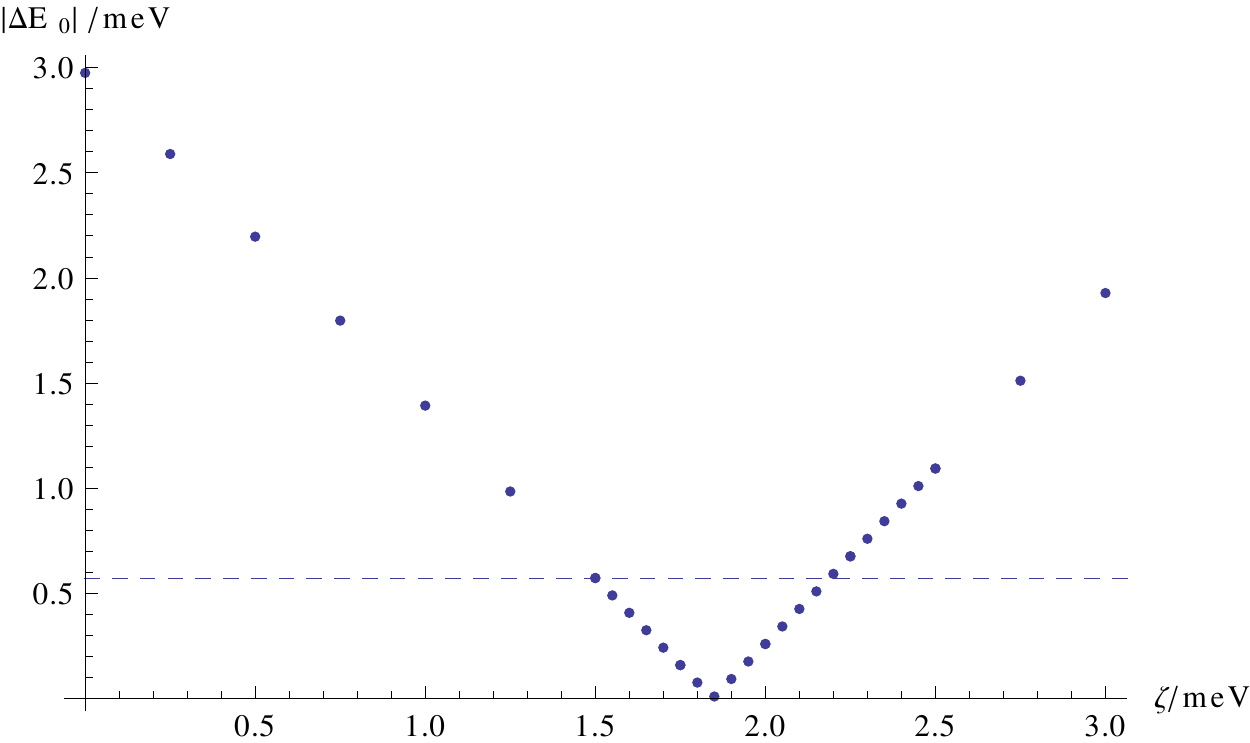}
 \caption{\label{pstrain}
 HH/LH splitting $\Delta E_0$ (in the absence of a quadrupole field) as a function of the strain-induced subband shift $\zeta$ for a QW thickness $W$ = 60 nm. Evidently, the (unwanted) HH/LH splitting $\Delta E_0$ can be tuned down to zero by a uniform strain in $z$ direction. The dashed line marks the value of the typical (wanted) HH/LH splitting $\Delta E = 0.57 {\rm meV}$ due to a quadrupole field as discussed in the main text.}
\end{figure}
Fig. \ref{pstrain} shows the zero-field splitting $\Delta E_0$ as function of strain, demonstrating that the confinement induced splitting can be tuned down to zero by means of uniaxial strain.

\begin{figure}
 \includegraphics[width=\columnwidth]{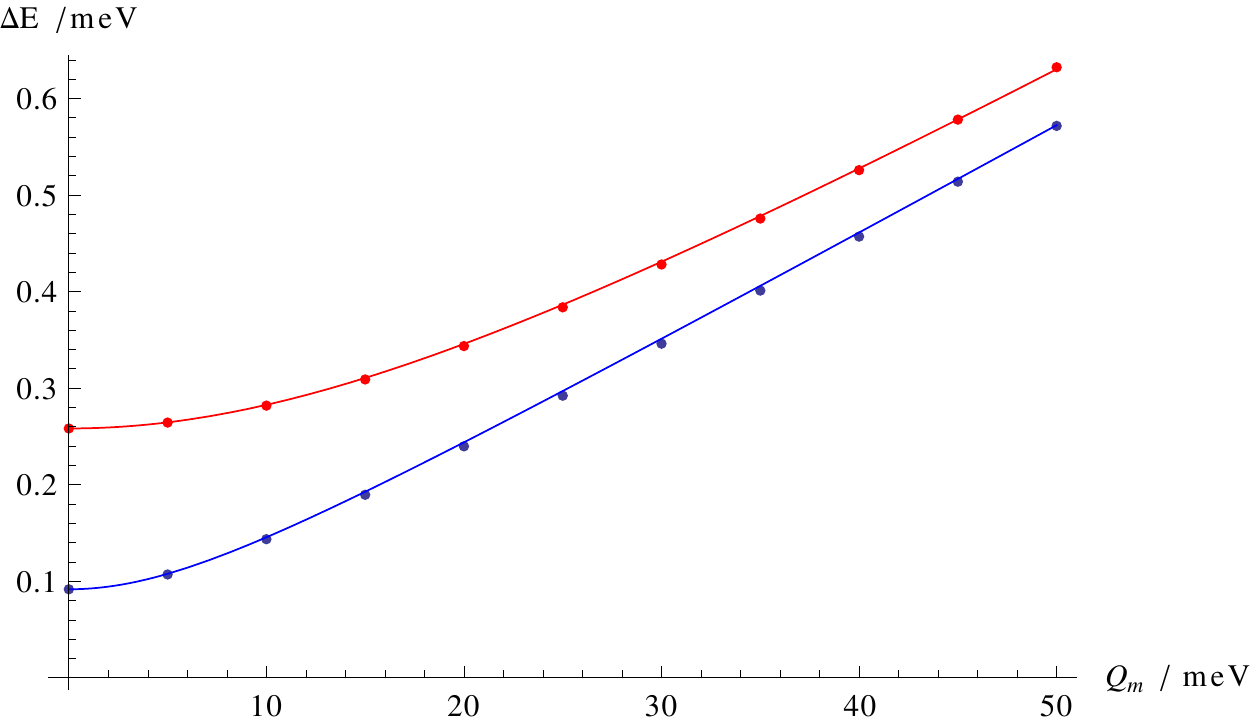}
 \caption{ \label{pfit}
 HH/LH splitting $\Delta E$ as function of the quadrupole potential $Q_m=\rm{max} (e\Phi_4)$ at $r = 50 nm$.
  The dots are numerical results and the continuous lines fits to the expected dispersion \eqref{dispeff} with the quadrupole parameter
  $\lvert \mathbf{x}\rvert = 0.00575 Q_m$ (red) and $\lvert \mathbf{x}\rvert = 0.00565 Q_m$ (blue).
  The full red line corresponds to a strain energy $\zeta = 2 \text{meV}$ and the full blue line to $\zeta = 1.9 \text{meV}$.
 }
\end{figure}
Fig. \ref{pfit} shows fits to the dispersion \eqref{dispeff} in order to obtain the quadrupole induced splitting $\lvert \mathbf{x}\rvert$.
A realistic quadrupole with a maximum potential $\left.e\Phi_4 \right |_{r=50nm}$~of $50\text{meV}$~gives a quadrupole induced splitting of $2 \lvert \mathbf{x}\rvert \approx$ 0.57 meV.

\subsection*{Stability of the quantum dot setup against perturbating potentials}
The aim of this section is to analyze the stability of the effective quadrupole Hamiltonian $H'$
against deviations from a perfect quadrupole potential with $l=2$. These deviations include external dipole fields and deviations from the
quadratic confinement and will be described as $ V(r,\phi)$ in the following.
The stability of $H'$ implies the stability of the quadrupole Hamiltonian $H(\mathcal{Q})$ since a change in the unwanted $\Delta E_0$ can be suppressed by adjusting the strain.

We consider the axial multipole expansion of the in-plane potential $V(r,\phi)$ given by a distribution of Coulomb charges $\rho(R,\phi')$.
The QD extension is small against the distance to the gates, i.e. $r \ll R$. We expand in the Legendre Polynomials $P_l$,
\begin{align}
\label{Vexpansion}
& V(r,\phi)  = \frac{e}{4 \pi \epsilon} \sum_{l=0}^{\infty} r^l \int_0^{2 \pi} d \phi' P_l(\cos (\phi-\phi'))\nonumber\\
 &\int_0^\infty dR \frac{1}{R^l} \rho(R,\phi')
\end{align}
We continue by expanding the $P_l$ as
\begin{equation}
\label{Plexpansion}
 r^l P_l (\cos (\phi-\phi')) = r^l \sum_{j=l,l-2,..} \alpha_{l,j} \cos(j (\phi-\phi')).
\end{equation}
For the  quadrupole symmetry  $V(r,\phi + \frac{\pi}{2}) = -V(r,\phi)$ and  upon inserting \eqref{Plexpansion} into \eqref{Vexpansion},
the nonzero coefficients $\alpha_{l,j}$ have $j=2,6,10,\dots$ and $j\le l$.
Similiarly, for the dipole symmetry $V(r,\phi + \pi ) = - V(r,\phi)$, the nonzero coefficients $\alpha_{l,j}$ fullfill $j=1,3,5,\dots$ and $j\le l$.
Table \ref{tabperturbations} shows how some characteristic terms in the expansion \eqref{Vexpansion} enter our model.
\begin{table}[h]
\begin{tabular}{ll|p{4.5cm}}
 $l=0$ &            &  Overall shift in energy that does not change $\Delta E$.  \\
 \hline
 $l=1$ & $r \cos \phi$  &  Shift of the center of the bound state assuming that quadrupole and confining potentials ( $\Phi_1 + \Phi_4$ ) are  quadratic in $r$. $\Delta E$ unchanged.  \\
 \hline
 $l=2$ &  $r^2$, $r^2 \cos 2\phi$     & Included in the model as  $\Phi_1 + \Phi_4$.    \\
 \hline
 $l=3$ & $r^3 P_3 = r^3 (\frac{3}{8} \cos \phi + \frac{5}{8} \cos 3\phi) $          &
          Lowest order that appears in dipole expansion and can induce quadratic Stark effect. \\
 \hline
 $l=4$ & $r^4 \cos 4\phi$    & Deviation from quadrupole symmetry by four equally charged gates.   \\
       & $r^4 \cos 2\phi$   & Allowed by quadrupole symmetry leading to the same effective Hamiltonian $H(\mathcal{Q})$ with $J=\frac{3}{2}$ but with the induced value
                                            $\Delta E$ only a few percent in comparison with $l=2$ term. Does not influence holonomy operations.
\\        & $r^4$              &  Correction to the confinement potential, which removes stability against the $l=1$ perturbation.
 \\  \hline
 $l=6$ &  $r^6 \cos 6\phi$   & Lowest order perturbation that appears in quadrupole expansion.
\end{tabular}
\caption{\label{tabperturbations}
Characteristic terms of the axial multipole expansion.
}
\end{table}

Let us now summarize the results included in Table I. The $l=0$ term induces an uninteresting energy shift.
The $l=1$ term could give rise to a linear or quadratic Stark effect.
However, in very good approximation, we may assume that GaAs and AlAs have inversion symmetry and can be described by a Luttinger Hamiltonian.
Since the Luttinger Hamiltonian $H_L$ is even under inversion, the lowest bound states
have even parity. This excludes the linear Stark effect by symmetry. Further, as long as we model both the confinement and the quadrupole potential as quadratic in $r$,
a linear potential will simply shift the center of the wave function.
Thus, the quadratic Stark effect also cannot change $\Delta E$.

For a numerical estimate of higher-$l$ terms, we model the gates by four Coulomb charges at $r=50 nm$.
We find that the $l=3$ and $l=4$ terms barely change $\Delta E$ even if the corresponding charge imbalance at the gates is highly overestimated as compared to a realistic experimental setup, meaning we have chosen them of the order of the quadrupole charges itself.
If quadrupole symmetry of the potential holds, the lowest perturbation term is $l=6$.
This term will change depending on the shape of the gates, but, since it contains a small parameter $(r^6/R^6)$, it is negligible.

\begin{figure}
 \includegraphics[width=\columnwidth]{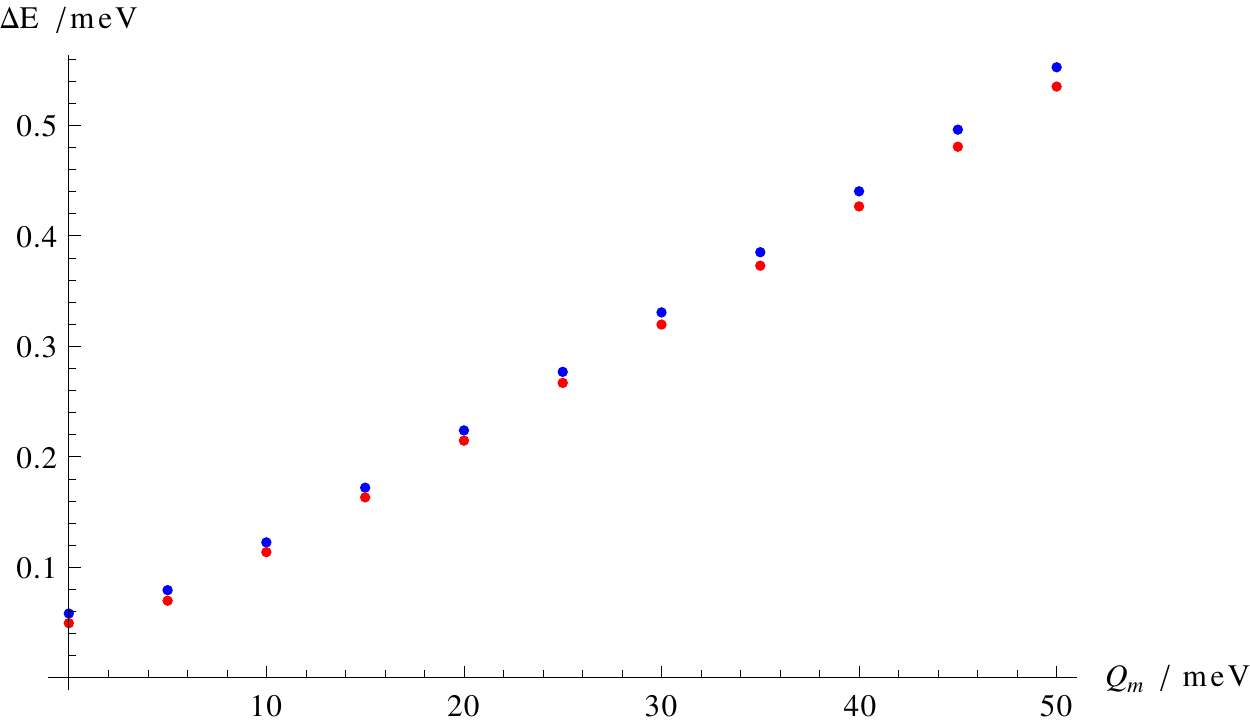}
 \caption{ \label{pfitvc4}
  Including a $r^4$ correction to the confinement ($\Phi_1'$ in the text) allows for the quadratic Stark effect by a homogeneous electric field.
  The plot shows the
 HH/LH splitting $\Delta E$ as function of the quadrupole potential $Q_m=\rm{max}\, (e\Phi_4$) at $r = 50 nm$ with $\zeta = 1.9 \text{meV}$ and $W = 60 nm$.
  Blue dots are without the dipole potential $\Phi_2$ while red dots include the $\Phi_2$, which corresponds to a charging ratio of 1/3 of a dipole vs. quadruple configuration.
  This ratio certainly overestimates the error that we expect in the experimental situation.
 }
\end{figure}

Finally, we note that the system is no longer robust against the quadratic Stark effect if the confinement potential behaves other than $r^2$.
We analyze this case in Fig. \ref{pfitvc4}, by changing the in-plane confinement to $e\Phi_1' = -0.15 \text{eV} (r/R_{max})^2 (1 + \frac{1}{3}\frac{r^2}{R_{max}^2})$.
A residual constant dipole field is modeled by an additional potential $e\Phi_2 = -0.025 \text{eV} \frac{r}{R_{max}} \cos(\phi - \pi/3)$ so that it is not aligned with the other potentials,
and corresponds to a dipole charging being $\frac{1}{3}$ of the quadrupole charging. This certainly overestimes the error expected in an experiment. Nevertheless, as can be seen in Fig. \ref{pfitvc4}, $\Delta E$ is barely affected by this perturbation.

Summarizing, we find that the effective Hamiltonian $H(\mathcal{Q})$ remains valid in good approximation. In all cases, the quadrupole splitting dominates the other (disturbing) contributions for realistic parameters.
\bibliographystyle{apsrev}
\bibliography{squad}
\end{document}